\documentclass[11pt,preprint]{aastex}
\usepackage{epsfig}
\usepackage{natbib}
\usepackage{graphicx}
\usepackage{slashbox}
\usepackage{multirow}
\usepackage{lscape}
\usepackage{mathrsfs,amssymb}
\usepackage{amsmath}
\newcommand       \Angstrom     {\,{\rm \AA}}

\newcommand       \cm           {\,{\rm cm}}

\newcommand       \eV           {\,{\rm eV}}

\newcommand       \K            {\,{\rm K}}

\newcommand       \s            {\,{\rm s}}
\newcommand       \sr           {\,{\rm sr}}

\newcommand       \GHz     {\,{\rm GHz}}

\newcommand       \mum          {\,{\rm \mu m}}

\newcommand       \Teff         {T_{\rm eff}}
\newcommand       \Tstar      {T_{\rm eff}}

\newcommand       \simali       {\sim\,}
\newcommand       \magni        {\,{\rm mag}}

\newcommand	  \NH          {N_{\rm H}}
\newcommand           \Ccnn       {\left[\frac{\rm C}{\rm H}\right]_{\rm CNN}}
\def    \Nb	{M}
\def    \bT	{{\bf T}}

\def    \abs       {{\rm abs}}

\countdef\decade=200
\advance\decade by \year
\countdef\hours=201
\advance\hours by \time
\divide\hours by 60
\countdef\mins=202
\advance\mins by \hours
\multiply\mins by 60
\multiply\hours by 100
\countdef\miltime=203
\advance\miltime by \hours
\advance\miltime by \time
\advance\miltime by -\mins
\def\today{\number\decade.\number\month.\number\day.\number\miltime}

\shorttitle{Cyanonaphthalenes in the Interstellar Medium}
\title{
\vspace*{-2.0em}
{\normalsize\rm Accepted for publication in
               {\it The Astrophysical Journal}.}\\
\vspace*{1.0em}
Infrared Emission of Specific Polycyclic Aromatic
Hydrocarbon Molecules: Cyanonaphthalenes 
\\{\small DRAFT: \today ~~}
}
\author{
Kaijun~Li\altaffilmark{1},
Aigen Li\altaffilmark{2},
X.J.~Yang\altaffilmark{3}, and
Taotao~Fang\altaffilmark{1}
}
\altaffiltext{1}{Department of Astronomy,
  Xiamen University, Xiamen, Fujian 361005, China;
  {\sf fangt@xmu.edu.cn}}
\altaffiltext{2} {Department of Physics and Astronomy,
                  University of Missouri,
                  Columbia, MO 65211, USA;
                  {\sf lia@missouri.edu}}
\altaffiltext{3}{Department of Physics,
                      Xiangtan University,
                      411105 Xiangtan, Hunan Province, China;
                      {\sf xjyang@xtu.edu.cn}}
\begin{document}

\begin{abstract}
The unidentified infrared emission (UIE) features 
at 3.3, 6.2, 7.7, 8.6, 11.3 and 12.7$\mum$ are ubiquitously
seen in a wide variety of astrophysical regions and 
commonly attributed to polycyclic aromatic hydrocarbon 
(PAH) molecules. However, the unambiguous identification
of any individual, specific PAH molecules has proven elusive
until very recently two isomers of cyanonapthalene,
which consists of two fused benzene rings 
and substitutes a nitrile (–-CN) group for a hydrogen atom,
were discovered in the Taurus Molecular Cloud 
based on their rotational transitions at radio frequencies.
To facilitate the {\it James Webb Space Telescope} (JWST)
to search for cyanonapthalenes in astrophysical regions,
we model the vibrational excitation of cyanonapthalenes
and calculate their infrared emission spectra
in a number of representative astrophysical regions.
The model emission spectra and intensities
will allow {\it JWST} to quantitatively determine
or place an upper limit on
the abundances of cyanonapthalenes.

\end{abstract}

\section{Introduction\label{sec:intro}}
The ``unidentified infrared emission'' (UIE) bands, 
a distinct set of spectral features at wavelengths 
of 3.3, 6.2, 7.7, 8.6, 11.3 and 12.7$\mum$, 
dominate the infrared (IR) spectra of many bright 
astronomical objects. They are ubiquitously seen in 
the interstellar medium (ISM) of our own galaxy and 
star-forming galaxies,
both near (e.g., the Local Group)
and far (e.g., distant galaxies at redshifts $z>4$), 
and account
for up to 20\% of their total infrared (IR) 
luminosity (see Li 2020).
The identification of the UIE bands is important as 
they are a useful probe of the cosmic star-formation history, 
and their carriers are an essential player in galactic evolution.
Although the exact nature of the UIE bands remains unknown,
a popular hypothesis is that they arise from
the C--H and C--C vibrational transitions of
polycyclic aromatic hydrocarbon (PAH) molecules 
(L\'eger \& Puget 1984, Allamandola et al.\ 1985, Tielens 2008).
%

\begin{figure}[h]
\centering
\includegraphics[height=12cm,width=12cm]{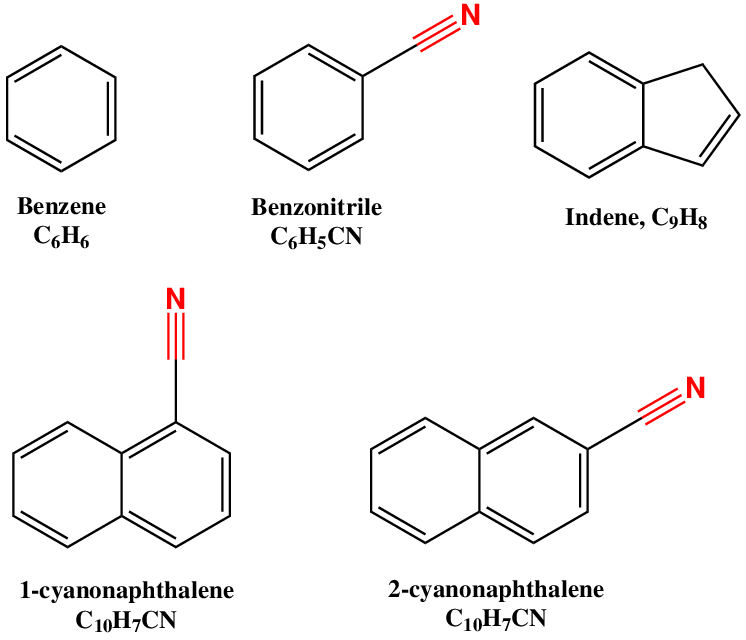}
\caption{\footnotesize
         \label{fig:specpah}
         Chemical structures of specific aromatic molecules
         identified in the ISM.
         }
\end{figure}

Nevertheless, despite the popularity of the PAH hypothesis,
unambiguous astronomical identification of
any individual, specific PAH molecules
has long proven elusive for nearly four decades
ever since the PAH model was first proposed in the 1980s.
However, recent years have witnessed breakthroughs
in this aspect. Based on radio observations
of the dark molecular cloud TMC-1
located within the Taurus Molecular Cloud
made by the 100\,m Green Bank Telescope (GBT)
in the frequency range of 8 to 34$\GHz$,
McGuire et al.\ (2021) reported the detection of
the rotational transitions of
1-cyanonaphthalene (1-CNN)
and 2-cyanonaphthalene (2-CNN),
two isomers of cyanonaphthalene (C$_{10}$H$_7$CN),
in which a nitrile/cyano (--CN) group replaces one of
the hydrogen atoms of naphthalene
(see Figure~\ref{fig:specpah}).
%
This is the first definitive identification
of a specific PAH molecule in space.
Similarly, Burkhardt et al.\ (2021) conducted
GBT radio observations of TMC-1 
in the frequency range of 2--12 and 18--34$\GHz$.
They detected indene (C$_9$H$_8$),
a pure hydrocarbon PAH species
composed of both a five- and six-membered ring
(see Figure~\ref{fig:specpah}).

Prior to these, benzene (C$_6$H$_6$) was detected
in CRL~618, a protoplanetary nebula,
through a single weak absorption feature
arising from its $\nu_4$ bending mode
at $\simali$14.85$\mum$ (Cernicharo et al.\ 2001).
Recently, benzene has also been detected by
the {\it James Webb Space Telescope} (JWST)
through the 14.85$\mum$ emission feature
in a protoplanetary disk around a low-mass star
(Tabone et al.\ 2023).
Benzonitrile (C$_6$H$_5$CN),
a single benzene ring with an attached CN group,
was discovered also with GBT by McGuire et al.\ (2018)
in TMC-1. This discovery was also achieved by
radio observations of its rotational lines
in the frequency range of 18 to 23$\GHz$.
By definition, benzene and benzonitrile
are aromatic but not PAH molecules.
For illustration, Figure~\ref{fig:specpah} lists
all these specific aromatic species
spectroscopically identified in space.\footnote{%
  Several small PAHs, including 
  naphthalene (C$_{10}$H$_{8}$), 
  phenanthrene (C$_{14}$H$_{10}$),
  perylene (C$_{16}$H$_{10}$),
  pyrene (C$_{20}$H$_{12}$), and
  their derivatives 
  have been found in the {\it Stardust} samples 
  collected from comet Wild\,2 
  (Sandford et al.\ 2006, Clemett et al.\ 2010),
  in the {\it Rosetta} samples collected from
  comet 67P/Churyumov-Gerasimenko
  (Schuhmann et al.\ 2019),
  in the {\it Hayabusa 2} samples collected
  from the carbonaceous asteroid Ryugu
  (Naraoka et al.\ 2023),
  and in interplanetary dust 
  particles possibly of cometary origin
  (Clemett et al.\ 1993).
  These PAH molecules were identified
  through mass spectrometry,
  not through vibrational or rotational spectroscopy.
  }
These aromatic molecules may be precursors to
more complex PAHs. The identification of these specific
aromatic molecules sheds light on the composition of
aromatic material within the ISM that will eventually be
incorporated into new stars and planets.
%



Nevertheless, identification of specific PAH molecules
through IR spectroscopy has so far been ineffective.
The UIE bands are attributed to collective emission
from many different PAH species and thus do not
allow one to fingerprint individual PAH molecules.
However, the IR spectra of different PAH molecules
are readily distinguishable in the laboratory
from one species to another.
Therefore, in principle, IR spectroscopy could
provide individual identification if the molecule 
is sufficiently abundant and the telescope instrument
is sufficiently sensitive.
With the advent of JWST,
this may become possible
due to its unprecedented sensitivity.
%
With an aim to offer spectroscopic guidance
for JWST to search for specific PAH molecules,
we perform a systematic exploration
of the theoretical IR emission spectra of various
specific PAH species expected in different
astrophysical environments.
In this work, we report the model IR emission
spectra for cyanonaphthalenes (i.e., 1- and 2-CNN),
the first PAH molecules ever detected in the ISM.
In \S\ref{sec:cabs} we synthesize  
the ultraviolet (UV) absorption cross sections
which determine how cyanonaphthalenes
absorb starlight in space. 
The IR absorption cross sections are also
discussed in \S\ref{sec:cabs} which determine
how cyanonaphthalenes emit in the IR.
The vibrational excitation and de-excitation
of cyanonaphthalenes are discussed in 
\S\ref{sec:model}.
We present in \S\ref{sec:irem} the model
IR emission spectra of cyanonaphthalenes.
The results are discussed in \S\ref{sec:discussion}
and summarized in \S\ref{sec:summary}.



%


\section{UV and IR Absorption Cross Sections}\label{sec:cabs}
\begin{figure}[h]
\centering
\includegraphics[height=12cm,width=12cm]{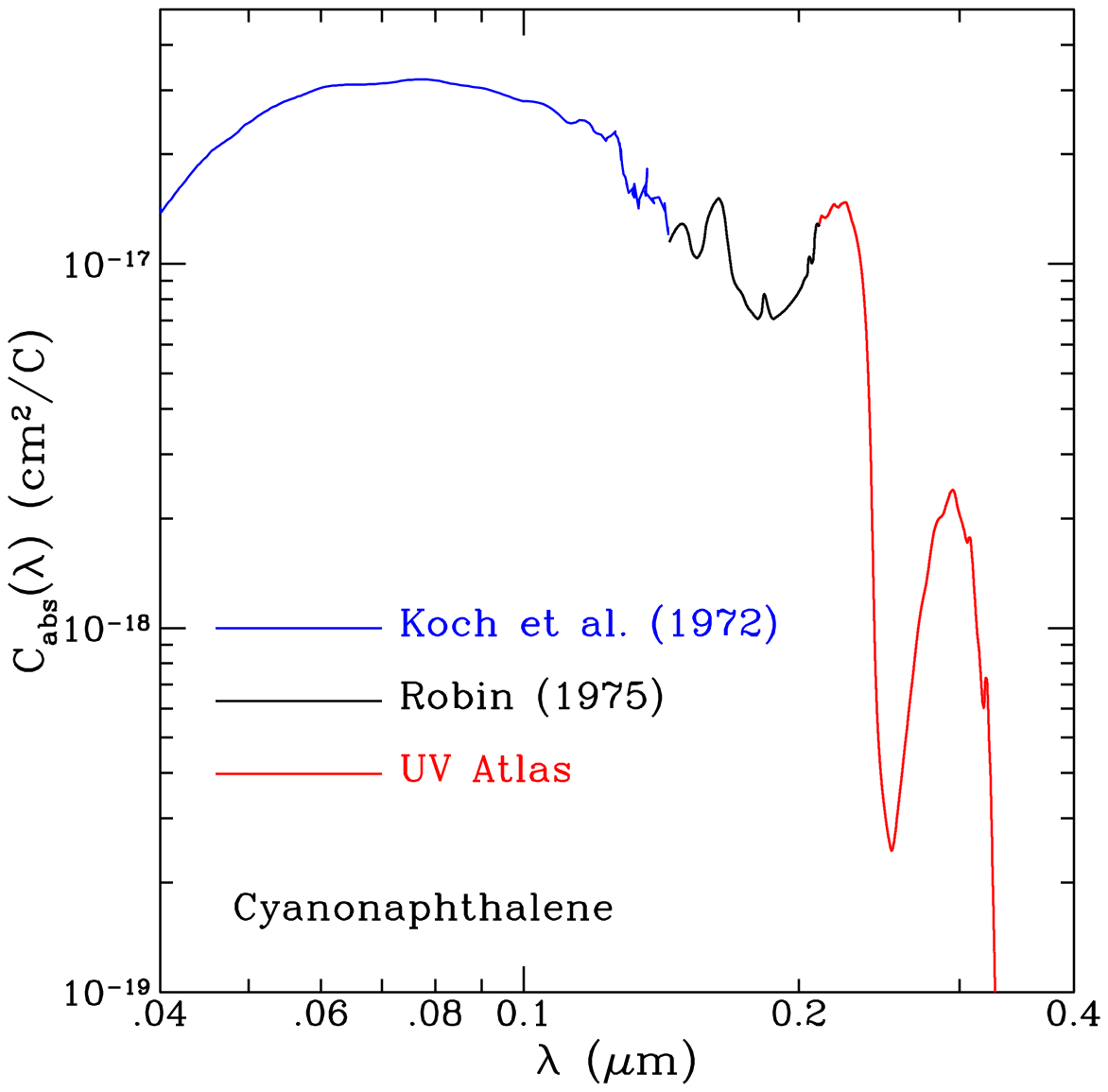}
\caption{\footnotesize
         \label{fig:cabs_uv}
         UV absorption cross sections (per C atom)
         of cyanonaphthalenes ``synthesized'' from
         the experimental data of 1-CNN over 
         $0.21\mum <\lambda<0.33\mum$
         (red line; Perkampus 1992),
         and of naphthalene over
         $0.14\mum <\lambda<0.21\mum$
         (black line;  Robin 1975)
         and over $0.04\mum <\lambda<0.14\mum$
         (blue line; Koch et al.\ 1972).
         }
\end{figure}

The UV absorption cross sections of cyanonaphthalenes
determine how they absorb starlight photons in space.
Unfortunately, the UV absorption has only been
measured for 1-CNN over the wavelength range of
$0.21\mum <\lambda<0.33\mum$ (Perkampus 1992).
Therefore, for $0.14\mum <\lambda<0.21\mum$ we
adopt the UV absorption of naphthalene (Robin 1975)
and for $0.04\mum <\lambda<0.14\mum$
we take the vacuum UV absorption
measured by Koch et al.\ (1972) for naphthalene.
Finally, as shown in Figure~\ref{fig:cabs_uv}, 
the UV absorption spectra of naphthalene are
smoothly scaled to join that of 1-CNN.
As cyanonaphthalenes are too small to absorb
appreciably in the visible, for $\lambda>0.33\mum$
we therefore extrapolate from that of Perkampus (1992)
at $\lambda<0.33\mum$.
Due to lack of experimental data, we do not distinguish
the UV absorption between 1-CNN and 2-CNN.

\begin{figure}[h]
\centering
\includegraphics[height=12cm,width=12cm]{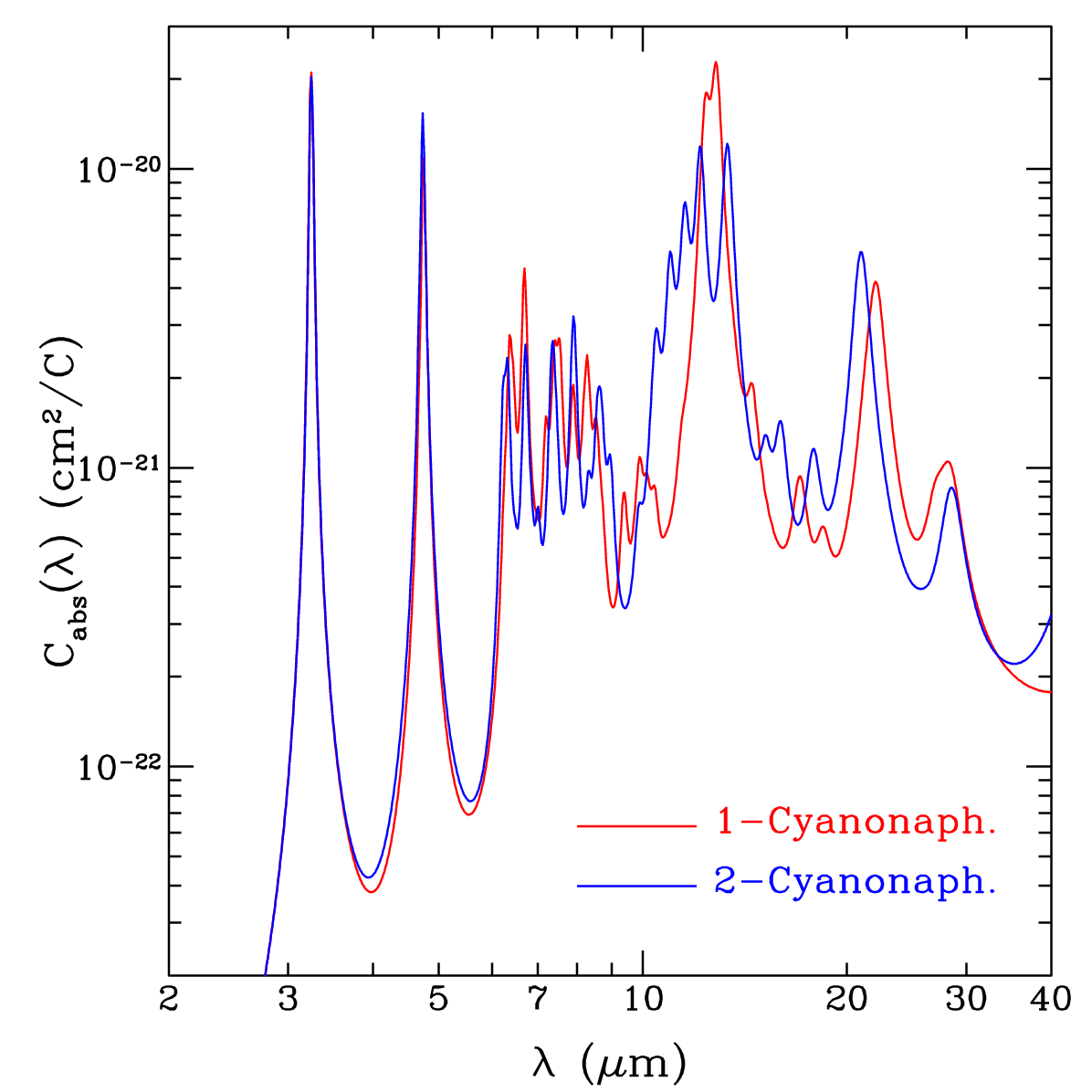}
\caption{\footnotesize
         \label{fig:cabs_ir}
         IR absorption cross sections (per C atom)
         of cyanonaphthalenes
         (red line: 1-CNN; blue line: 2-CNN)
         calculated from the vibrational frequences
         and intensities computed by
         Bauschlicher  (1998)
         using B3LYP/4-31G.
         Each vibrational line is assigned
         a width of 30$\cm^{-1}$, consistent with
         the natural line width expected from
         a vibrationally excited PAH molecule 
         (see Allamandola et al.\ 1999).
         }
\end{figure}

In the IR, Bauschlicher (1998) have computed
the vibrational frequencies and intensities of
1-CNN and 2-CNN, using the B3LYP density
functional theory in conjunction with
the 4-31G basis set. We take the vibrational
frequencies and intensities of Bauschlicher (1998)
which are available from
the {\it NASA Ames PAH IR Spectroscopic Database}
(Boersma et al.\ 2014, Bauschlicher et al.\ 2018, 
Mattioda et al.\ 2020).
We represent each vibrational line by
a Drude function,\footnote{%
  If we approximate the vibrational modes of PAHs
  as harmonic oscillators,
  the absorption cross sections
  are then expected to be Drude functions
  (see Li 2009).
}
characterized by the peak wavelength and intensity
of the vibrational transition.
In addition, we assign a width of 30$\cm^{-1}$
for each line, consistent with the natural line width
expected from a vibrationally excited PAH molecule 
(see Allamandola et al.\ 1999).
The resulting IR absorption sections of
1-CNN and 2-CNN are shown in
Figure~\ref{fig:cabs_ir}.
It is remarkable that the IR vibrational spectra
of 1-CNN and 2-CNN are quite similar
at $\lambda<10\mum$ and even at 
$\lambda>10\mum$ they do not differ
substantially. This will be further discussed
in \S\ref{sec:irem}.

\section{Vibrational Excitation of
           Cyanonaphthalenes}\label{sec:model}
Cyanonaphthalenes only have 19 atoms and
51 vibrational degrees of freedom. 
Upon absorption of an UV stellar photon,
cyanonaphthalenes will undergo stochastic
heating since their energy contents are often
smaller than the energy of a single stellar photon.  
We model the stochastic heating 
of cyanonaphthalenes by employing
the ``exact-statistical'' method
of Draine \& Li (2001).
We characterize the state of
a cyanonaphthalene molecule
(i.e., 1-CNN or 2-CNN)
by its vibrational energy $E$, 
and group its energy levels into 
$(\Nb+1)$ ``bins'',
where the $j$-th bin ($j$\,=\,0,\,...,\,$\Nb$) 
is $[E_{j,\min},E_{j,\max})$, 
with representative energy 
$E_j$\,$\equiv$\,$(E_{j,\min}$+$E_{j,\max})$/2,
and width 
$\Delta E_j$\,$\equiv$\,$(E_{j,\max}$--$E_{j,\min})$.
Let $P_j$ be the probability of finding 
1-CNN (or 2-CNN) in bin $j$ with energy $E_j$.
The probability vector $P_j$ evolves 
according to
\begin{equation}
dP_i/dt = \sum_{j\neq i} \bT_{ij} P_j 
- \sum_{j\neq i} \bT_{ji}P_i ~~,~~ i\,=\,0,...,\Nb ~~,
\end{equation}
where the transition matrix element $\bT_{ij}$ is
the probability per unit time for
1-CNN (or 2-CNN) in bin $j$ 
to make a transition to one of the levels in bin $i$. 
We solve the steady state equations
\begin{equation}\label{eq:steadystate}
\sum_{j\neq i} \bT_{ij} P_j
= \sum_{j\neq i} \bT_{ji}P_i ~~,~~ i\,=\,0,...,\Nb ~~
\end{equation}
to obtain the $\Nb$+1 elements of $P_j$,
and then calculate the resulting IR emission spectrum
(see eq.\,55 of Draine \& Li 2001).

In calculating the state-to-state transition rates
$\bT_{ji}$ for transitions $i$$\rightarrow$$j$,
we distinguish the excitation rates $\bT_{ul}$ 
(from $l$ to $u$, $l$\,$<$\,$u$) 
from the deexcitation rates $\bT_{lu}$ 
(from $u$ to $l$, $l$\,$<$\,$u$).
For a given starlight energy density $u_E$,
the rates for upward transitions $l$$\rightarrow$$u$ 
(i.e., the excitation rates)
are just the photon absorption rates:
\begin{equation}\label{eq:Tul}
\bT_{ul} \approx C_{\abs}(E)\,c\,u_E \Delta E_u/(E_u-E_l) ~~.
\end{equation}
The rates for downward transitions $u$$\rightarrow$$l$
(i.e., the deexcitation rates)
can be determined from the detailed balance analysis
of the Einstein $A$ coefficient:
\begin{equation}\label{eq:Tlu}
\bT_{lu} \approx \frac{8\pi}{h^3c^2} \frac{g_l}{g_u}
\frac{\Delta E_u}{E_u-E_l} E^3 \times C_{\abs}(E) 
\left[1+\frac{h^3c^3}{8\pi E^3}u_E\right] ~~,
\end{equation}
where $h$ is the Planck constant,
and the degeneracies $g_u$ and $g_l$ 
are the numbers of energy states 
in bins $u$ and $l$, respectively:
\begin{equation}
\label{eq:gj}
g_j \equiv N(E_{j,\max})-N(E_{j,\min})
\approx \left(dN/dE\right)_{E_j} \Delta E_j ~~,
\end{equation}
where $\left(dN/dE\right)_{E_j}$
is the vibrational density of states 
of 1-CNN (or 2-CNN) at internal energy $E_j$.
For a cyanonaphthalene molecule,
if we know the frequencies of 
all its 51 vibrational modes,
we can employ the Beyer-Swinehart numerical algorithm
(Beyer \& Swinehart 1973, Stein \& Rabinovitch 1973)
to calculate the vibrational density of states 
$\left(dN/dE\right)_{E_j}$
and therefore the degeneracies $g_j$ 
for each vibrational energy bin.
Meanwhile, if we know the oscillator strength
of each vibrational mode, 
we can obtain the IR absorption
cross section $C_{\rm abs}(E)$
by summing up all the vibrational transitions
with each approximated as a Drude profile.
Without a prior knowledge of the width
of each vibrational transition, 
we assign a width of 30$\cm^{-1}$, 
consistent with the natural line width 
expected from free-flying molecule 
(see Allamandola et al.\ 1999).
This natural line width arises from 
intramolecular vibrational energy redistribution. 

As mentioned earlier (see \S\ref{sec:cabs}),
Bauschlicher (1998) have computed
with B3LYP/4-31G the frequency
and intensity for each of the 51 vibrational
transitions of 1-CNN and 2-CNN.
Therefore, we can calculate the
vibrational density of states 
and the degeneracy $g_j$
for each vibrational energy bin
as well as the IR absorption cross section
$C_{\rm abs}(E)$
by adopting the vibrational frequencies
and intensities of 1-CNN and 2-CNN
computed by Bauschlicher (1998).
For a given astrophysical environment
characterized by its starlight energy density $u_E$,
we first calculate the excitation rates $\bT_{ul}$ 
(from $l$ to $u$, $l$\,$<$\,$u$)
according to eq.\,\ref{eq:Tul},
using the UV absorption cross setions
described in \S\ref{sec:cabs}.
We then calculate the deexcitation rates
$\bT_{lu}$ (from $u$ to $l$, $u$\,$>$\,$l$)
according to eq.\,\ref{eq:Tlu},
using the degeneracy $g_j$ and
the IR absorption cross section $C_{\rm abs}(E)$
derived from the vibrational frequencies
and intensities of Bauschlicher (1998).
With the state-to-state 
transition rates $\bT_{ji}$ determined,
we solve the steady-state probability 
evolution equation (see eq.\,\ref{eq:steadystate})
to obtain the steady-state energy probability 
distribution $P_j$ 
and finally calculate 
the resulting IR emission spectrum,
according to eq.\,55 of Draine \& Li (2001).
For computational convenience, 
we consider 500 energy bins
(i.e., $M=500$).

\section{Model Emission Spectra}\label{sec:irem}
\begin{figure*}
\begin{center}
\hspace{-1cm}
\begin{minipage}[t]{0.4\textwidth}
\resizebox{8.5cm}{7.5cm}{\includegraphics[clip]{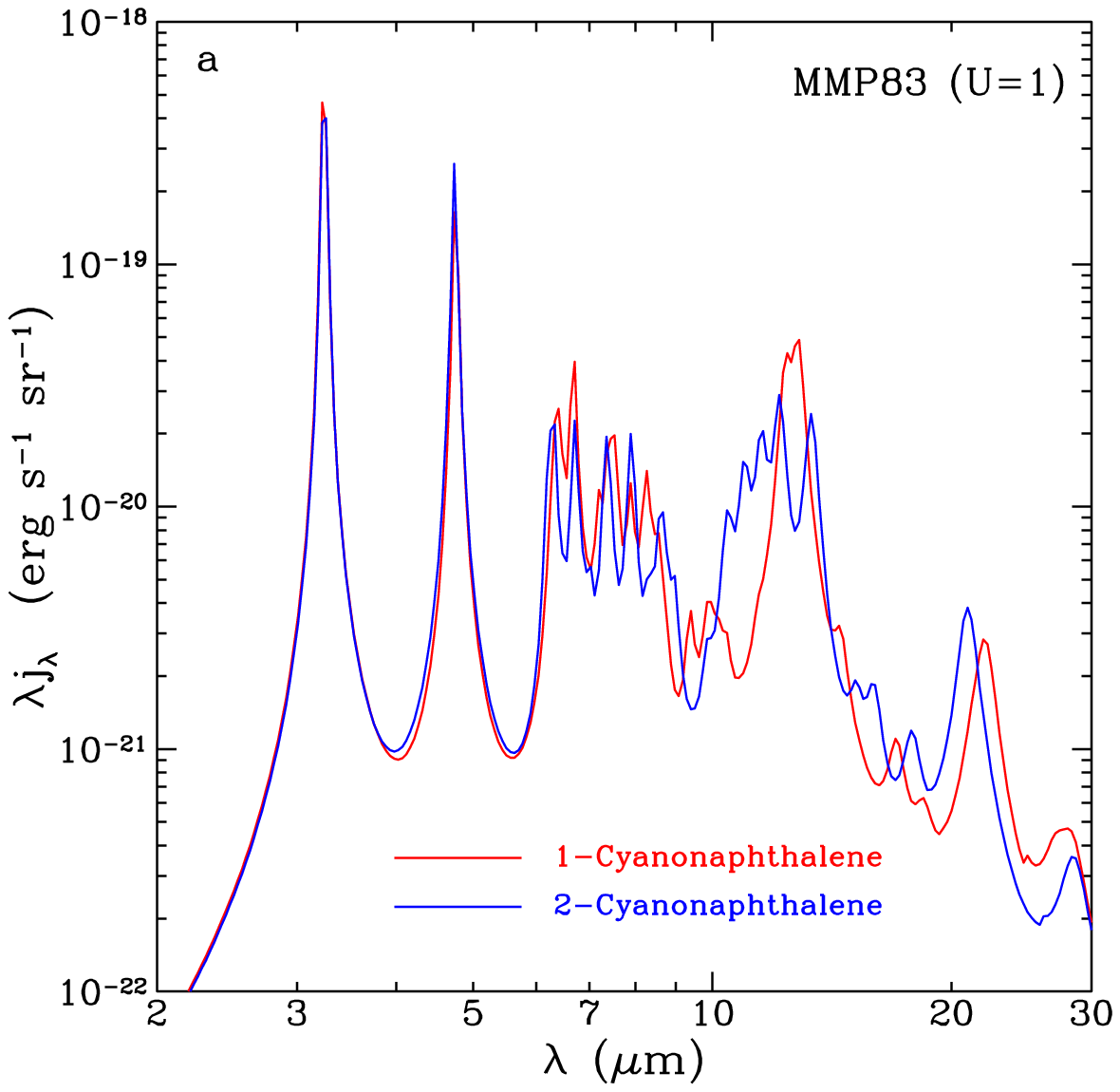}}\vspace{-0.5cm}
\end{minipage}
\hspace{2cm}
\begin{minipage}[t]{0.4\textwidth}
\resizebox{8.5cm}{7.5cm}{\includegraphics[clip]{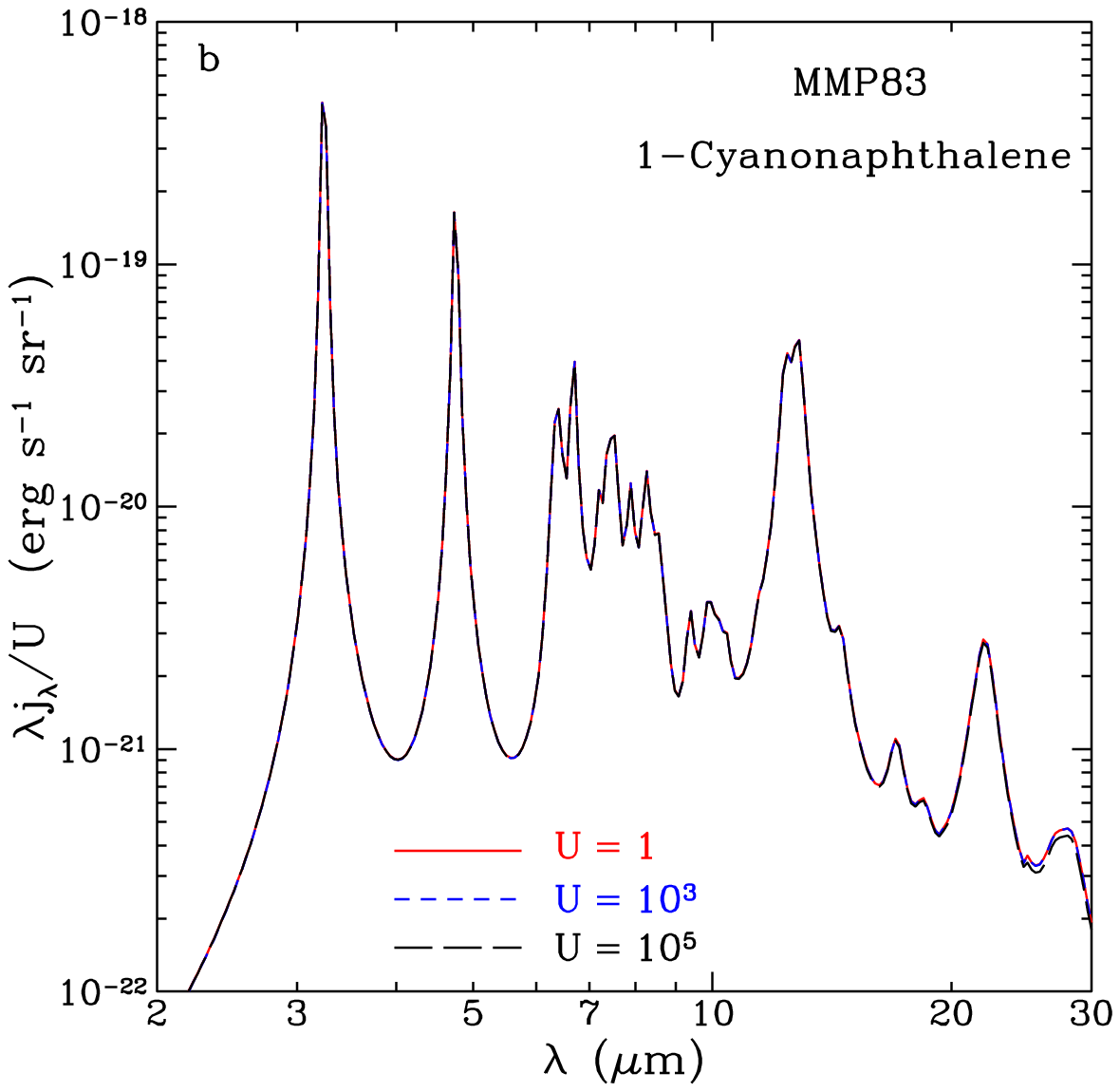}}\vspace{-0.5cm}
\end{minipage}
\end{center}
\caption{\footnotesize
         \label{fig:mmp83}
         Left panel (a): IR emission spectra of 1-CNN (red line)
         and 2-CNN (blue line) illuminated by the MMP83
         ($U$\,=\,1) interstellar radiation field.
         Right panel (b): IR emission spectra of 1-CNN
         illuminated by the MMP83 radiation field
         with different intensities
         (red solid line: $U$\,=\,1;
         blue shot-dashed line: $U$\,=\,1,000;
         black long-dashed line: $U$\,=\,10$^5$).
         }
\end{figure*}

We first consider cyanonaphthalenes in the diffuse ISM 
excited by the solar neighborhood interstellar radiation
field of Mathis et al.\ (1983; hereafter MMP83).
Figure~\ref{fig:mmp83}a shows the IR emissivities
(erg$\s^{-1}\sr^{-1}\cm^{-1}$) per molecule
for 1-CNN and 2-CNN.
A first glance of Figure~\ref{fig:mmp83}a
reveals that the IR emission spectra of 1-CNN
and 2-CNN are rather similar.
Both exhibit a prominent C--H stretching band
at 3.26$\mum$, a prominent C--N stretching band
at 4.69$\mum$, and a broad complex
at $\simali$6--9$\mum$ consisting of
a number of sub-features
arising from C--C stretching
and C--H in-plane bending vibrations.
The C--H out-of-plane bending bands
at $\simali$10.5--13.5$\mum$ are
also similar, except for 1-CNN they coalesce
into a broad band at 12.7$\mum$,
while for 2-CNN several sub-features
remain to be noticeable.
The C-C-C skeletal bending band occurs at
a longer wavelength for 1-CNN ($\simali$22$\mum$)
compared to 2-CNN ($\simali$21$\mum$),
otherwise they are also similar.
Because of the close similarity between
the emission spectra of 1-CNN and 2-CNN,
in the following we will only consider 1-CNN.

To verify the single-photon heating nature of
cyanonaphthalenes, we calculate the IR emission
of 1-CNN excited by enhanced MMP83-type radiation
fields. Figure~\ref{fig:mmp83}b shows the IR emission
spectra of 1-CNN excited by the MMP83 radiation field,
but enhanced by a factor of $U$\,=\,1000 and 10$^5$
($U$\,=\,1 corresponds to the MMP83 radiation field).
As illustrated in Figure~\ref{fig:mmp83}b, 
when scaled by $U$, the IR emission spectra are identical.
This justifies the stochastic heating treatment
of the vibrational excitation of cyanonaphthalenes.
In the single-photon heating regime, the resulting
IR emission is only dependent on the energy of
the illuminating starlight photons,
not on the starlight intensity.  

\begin{figure*}
\begin{center}
\hspace{-1cm}
\begin{minipage}[t]{0.4\textwidth}
\resizebox{8.0cm}{7.5cm}{\includegraphics[clip]{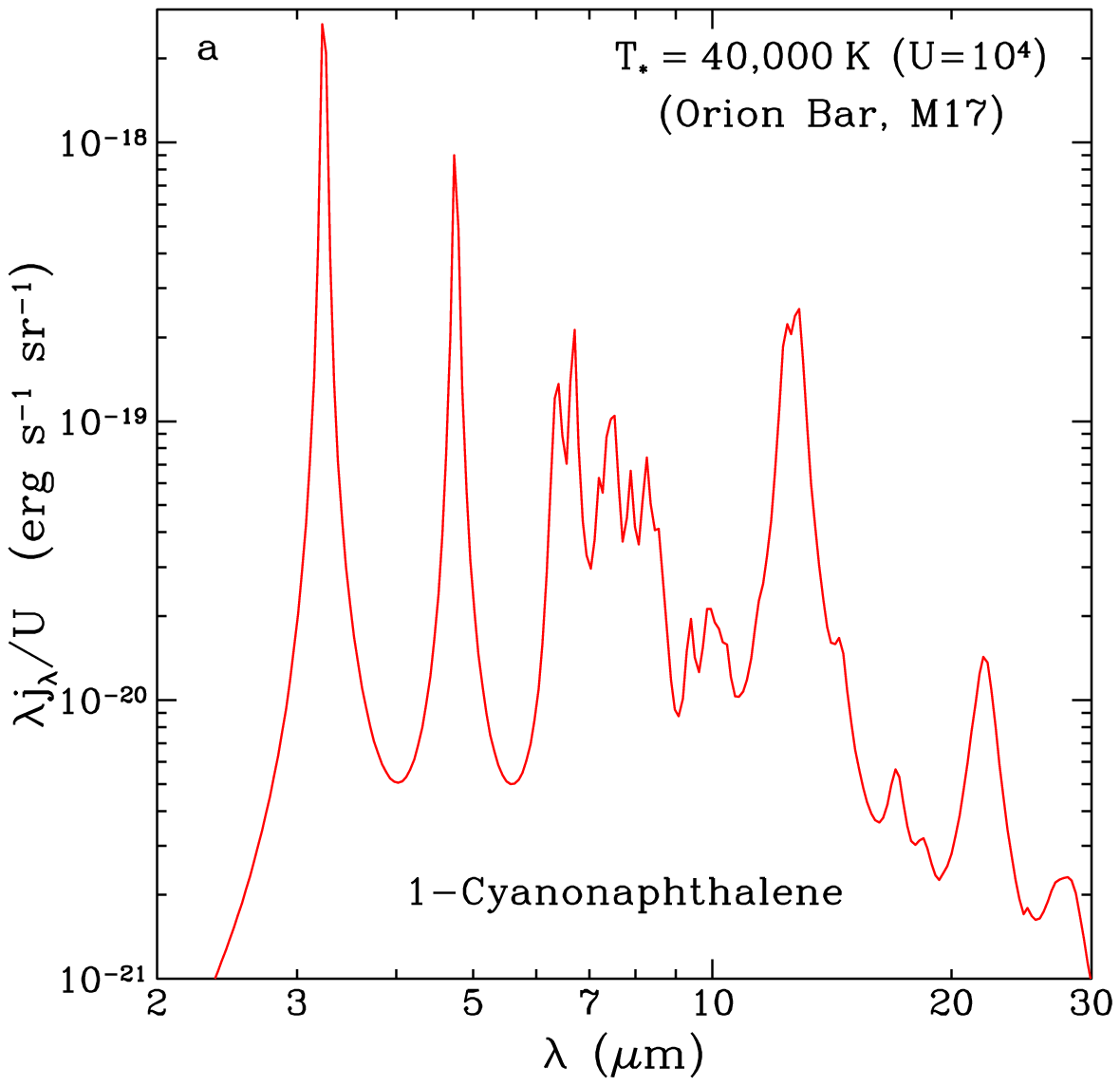}}\vspace{0.1cm}
\resizebox{8.0cm}{7.5cm}{\includegraphics[clip]{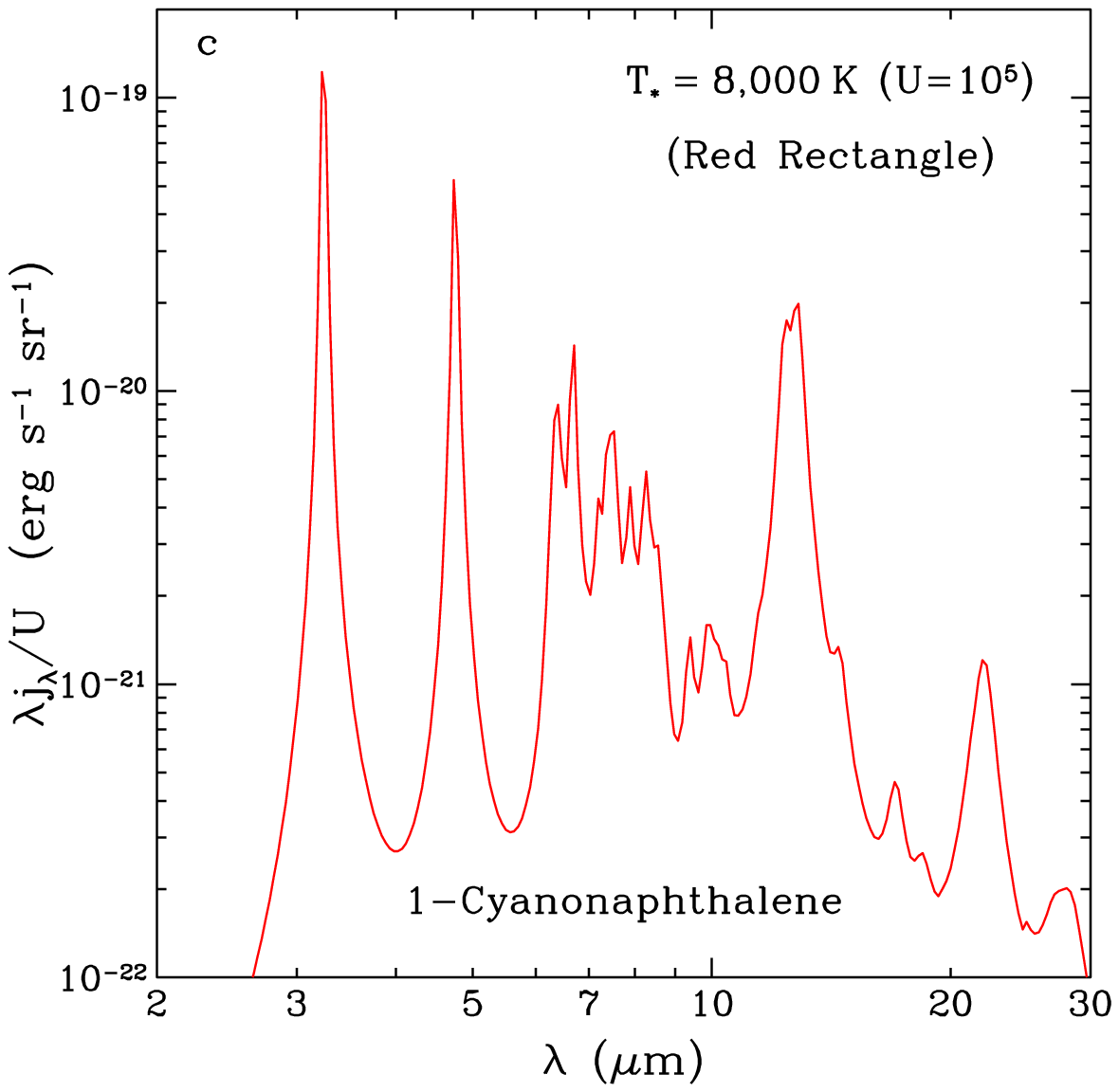}}\vspace{-0.5cm}
\end{minipage}
\hspace{2cm}
\begin{minipage}[t]{0.4\textwidth}
\resizebox{8.0cm}{7.5cm}{\includegraphics[clip]{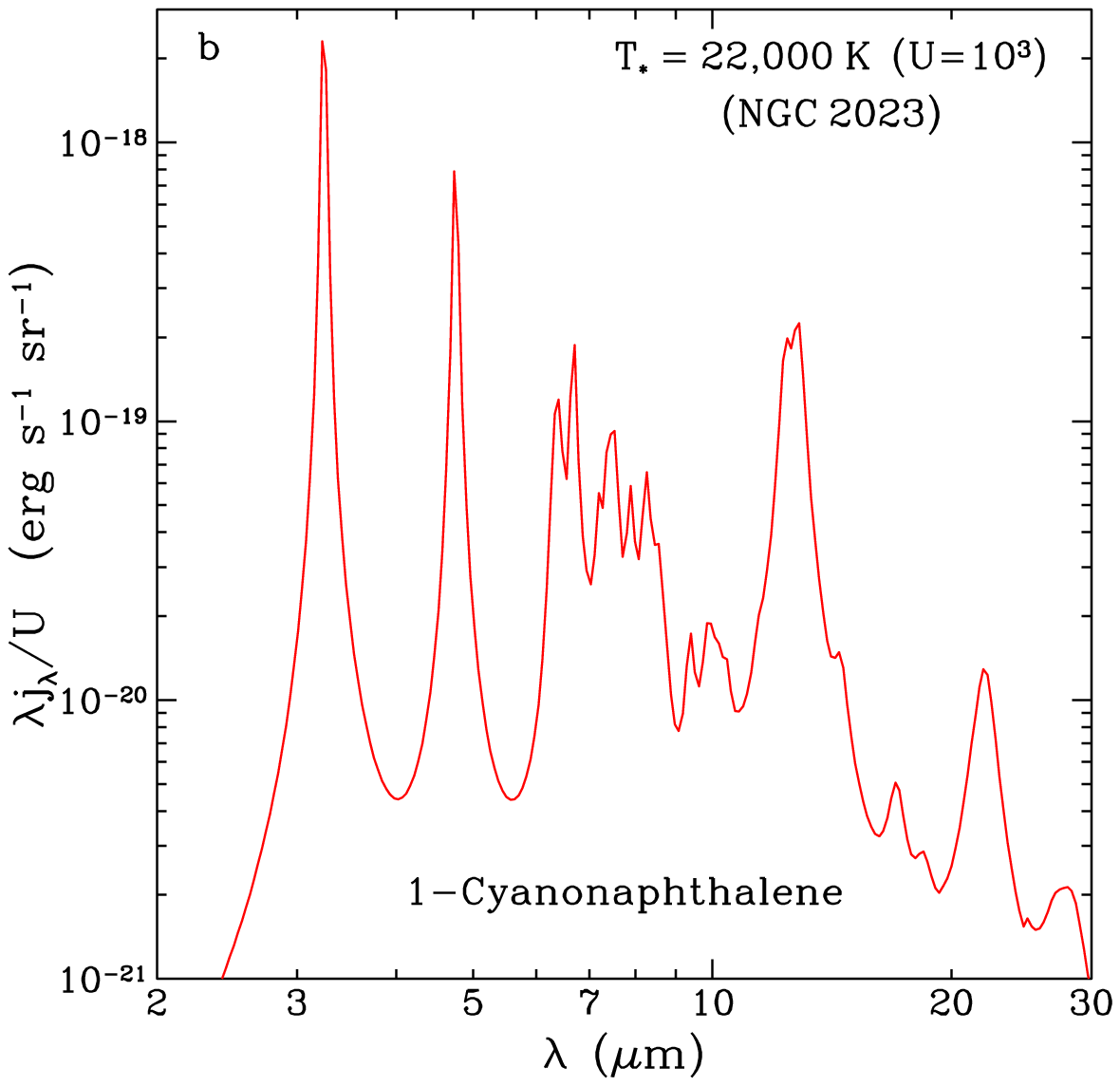}}\vspace{0.1cm}
\resizebox{8.0cm}{7.5cm}{\includegraphics[clip]{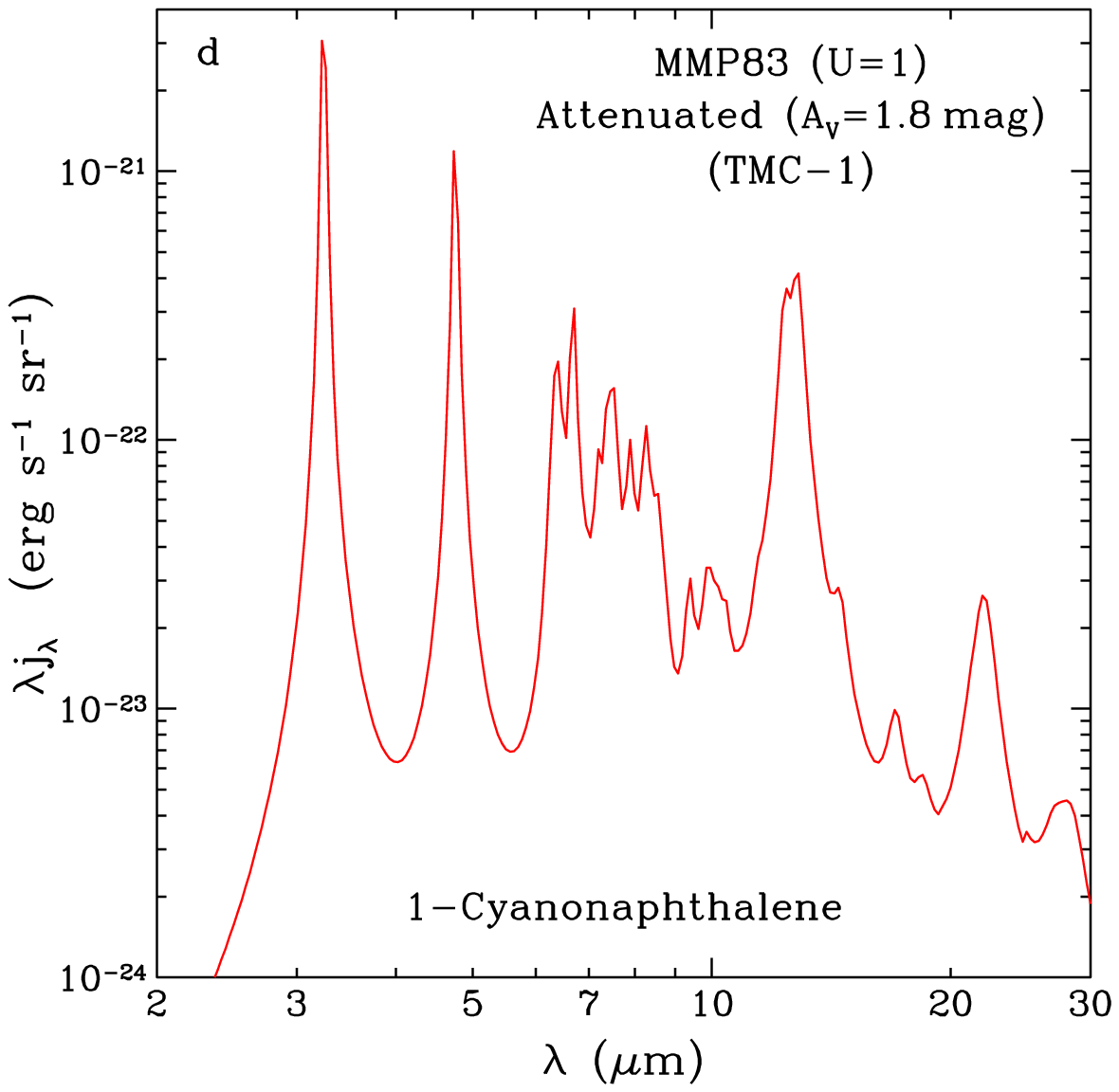}}\vspace{-0.5cm}
\end{minipage}
\end{center}
\caption{\footnotesize
         \label{fig:Teff}
         IR emission spectra of 1-CNN 
         illuminated by radiation fields
         of different starlight spectra:
         $\Teff$\,=\,40,000$\K$
         and $U$\,=\,10$^4$ (top left panel, a),
         $\Teff$\,=\,22,000$\K$
         and $U$\,=\,10$^4$ (top right panel, b),
         $\Teff$\,=\,8,000$\K$
         and $U$\,=\,10$^5$ (bottom left panel, c),
         and MMP83 radiation field
         attenuated by a visual extinction
         of $A_V=1.8\magni$ (bottom right panel, d).
         }
\end{figure*}

We also examine how the IR emission of cyanonaphthalenes
depends on the starlight spectrum (i.e., the ``hardness'' of
the exciting radiation field). To this end, we consider 1-CNN
exposed to stars with different effective temperatures:
$\Teff$\,=\,40,000, 22,000, 8,000$\K$.
The starlight spectra are approximated by
the Kurucz model atmospheric spectra.
Figure~\ref{fig:Teff}a shows the IR emission of 1-CNN
illuminated by stars of $\Teff$\,=\,40,000$\K$
(like O6V stars) with an intensity of $U$\,=\,10$^4$,
where $U$ is defined as
\vspace{-3mm}
\begin{equation}
U = \frac{\int_{1\mu {\rm m}}^{912{\rm \Angstrom}}
               4\pi J_\star(\lambda,\Tstar)\,d\lambda}
       {\int_{1\mu {\rm m}}^{912{\rm \Angstrom}}
               4\pi J_{\rm ISRF}(\lambda)\,d\lambda} ~~,
\end{equation}
where $J_{\rm ISRF}(\lambda)$ is 
the MMP83 radiation intensity, and
$J_\star(\lambda, \Tstar)$ is the intensity
of starlight approximated by the Kurucz model
atmospheric spectrum.
Such a starlight spectrum and intensity
resemble that of the Orion Bar
photodissociation region
and the M17 star-forming region.
A comparison of Figure~\ref{fig:Teff}a 
with  Figure~\ref{fig:mmp83}b reveals
that the IR emission spectra are almost
identical for 1-CNN excited by the MMP83 field
and by stars of $\Teff$\,=\,40,000$\K$,
except the emissivity level is higher for
the latter since the mean absorbed photon
energy is higher for the latter.

Figure~\ref{fig:Teff}b shows the IR emission of 1-CNN
illuminated by stars of $\Teff$\,=\,22,000$\K$
(like B1.5V stars) with an intensity of $U$\,=\,10$^3$.
This would be the case if 1-CNN is located
in the reflection nebula NGC\,2023.
If scaled by $U$ and the mean absorbed photon energy,
the IR emission spectrum for $\Teff$\,=\,22,000$\K$
is essentially identical to that for $\Teff$\,=\,40,000$\K$.

Figure~\ref{fig:Teff}c shows the IR emission of 1-CNN
illuminated by stars of $\Teff$\,=\,8,000$\K$
(like A5V stars) with an intensity of $U$\,=\,10$^5$.
The Red Rectangle protoplanetary nebula is illuminated
by such a starlight spectrum and intensity.\footnote{%
  It is worth noting that Witt et al.\ (2009) argued that
  the Red Rectangle protoplanetary nebula is illuminated
  by a binary, and the UV radiation in this nebula does not
  come from HD~44179 of which $\Teff\approx8,000\K$,
  but rather from the accretion disk surrounding the secondary
  star in this binary, with $\Teff$ in the range of
  $\simali$17,000--25,000$\K$.
  }
Cyanonaphthalenes in the Red Rectangle
would be excited by stellar photons
much softer than that in the Orion Bar.
It is therefore not unexpected that,
as illustrated in Figure~\ref{fig:Teff}c,
1-CNN in the Red Rectangle would emit
somewhat more at longer wavelengths
compared to that in the Orion Bar.
Nevertheless, the overall spectral shape
does not differ considerably between
that of the Red Rectangle and
that of the Orion Bar.

Finally, we also consider the TMC-1 molecular cloud
where cyanonaphthalenes were first detected through
their rotational lines. The TMC-1 cloud is externally
illuminated by the general interstellar radiation field
and has a total visual extinction of $\simali$3.6$\magni$
(Whittet et al.\ 2004). Therefore, to calculate the IR
emission of cyanonaphthalenes in the TMC-1 cloud,
we take the MMP83 interstellar radiation field,
but attenuated by dust extinction
with $A_V$\,=\,1.8$\magni$
(from the cloud surface to the cloud core)
and a wavelength-dependence $A_\lambda/A_V$
like that of the Galactic average extinction curve
of $R_V=3.1$, where $R_V$ is the total-to-selective
extinction ratio (see Cardelli et al.\ 1989),
i.e., cyanonaphthalenes in the TMC-1 cloud
are assumed to be excited by starlight
with an intensity of
$J_{\rm ISRF}(\lambda)\times
\exp\{-\left(A_V/1.086\right)
\times\left(A_\lambda/A_V\right)\}$.
Figure~\ref{fig:Teff}d shows the IR emission
of 1-CNN calculated for the TMC-1 cloud.
Compared to that excited by the MMP83
radiation field (see Figure~\ref{fig:mmp83}b),
the overall IR emission spectrum of 1-CNN
expected in the TMC-1 cloud is closely similar,
except the emissivity level is appreciably reduced
and the 3.26$\mum$ C--H stretch and 4.69$\mum$
C--N stretch emit slightly less. 

\section{Discussion}\label{sec:discussion}
\begin{figure}[h]
\centering
\includegraphics[height=12cm,width=12cm]{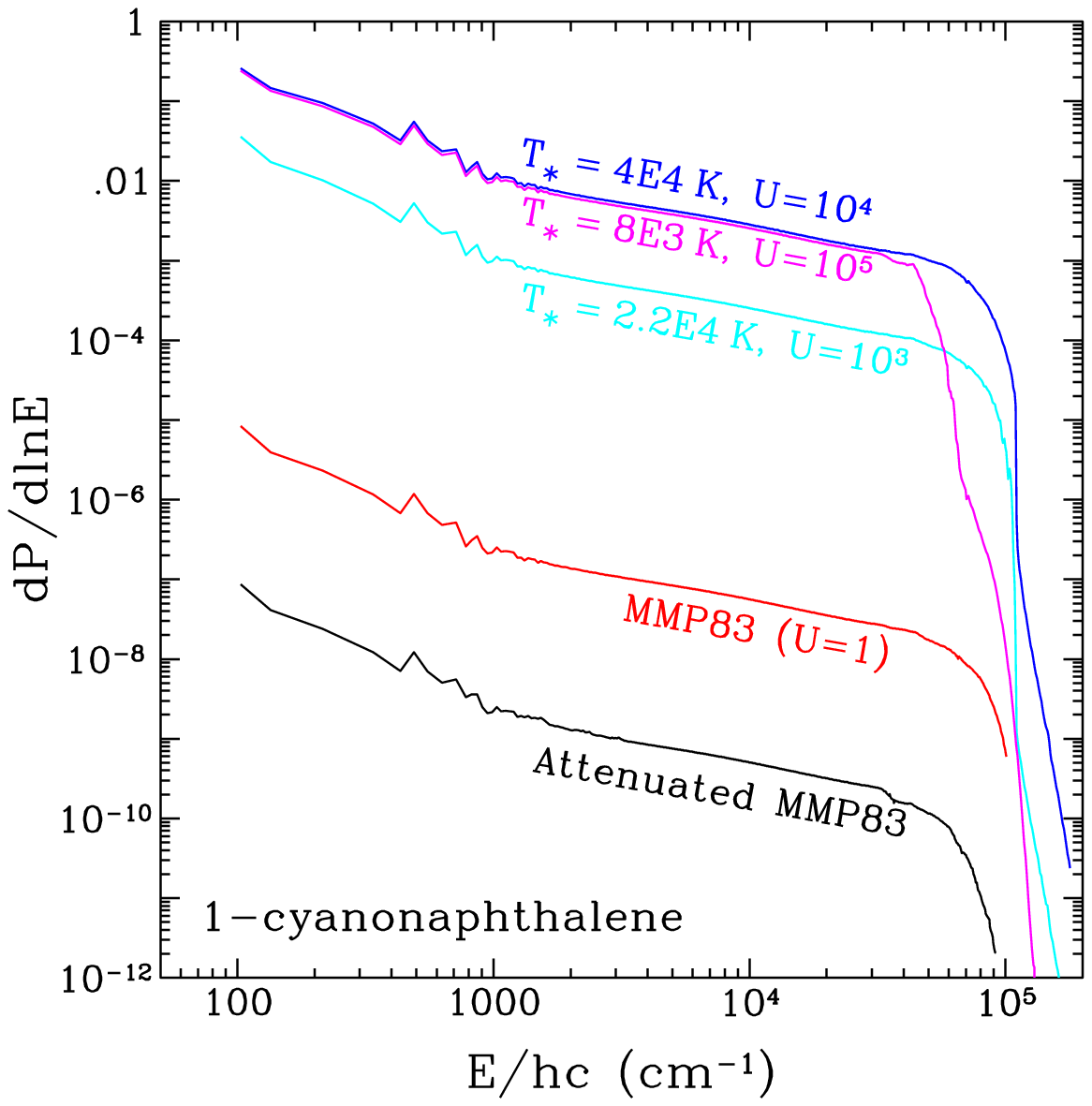}
\caption{\footnotesize
  \label{fig:dPdlnE}
        Vibrational energy probability distribution 
        functions for the excited vibrational states
        of 1-CNN, exposed to radiation fields of
        different intensities ($U$) and different
        spectral shapes:
        MMP83 (red),
        attenuated MMP83 (black), and stars
        of different effective temperatures
        (blue: $\Teff=4\times10^4\K$ and $U=10^4$,
        cyan: $\Teff=2.2\times10^4\K$ and $U=10^3$, and
        magenta: $\Teff=8\times10^3\K$ and $U=10^5$).
         }
\end{figure}

It is gratifying that, as illustrated in
Figure~\ref{fig:Teff}a--d, the IR emission spectra
of cyanonaphthalenes are not sensitive to
the illuminating starlight spectrum.
This is because, as shown in Figure~\ref{fig:dPdlnE},
due to their small heat contents,
the energy probability distribution functions of
cyanonaphthalenes closely resemble each
other, upon excited by starlight of different
intensities and different spectral shapes.
%

While observationally the identification of
cyanonaphthalenes through the $\simali$6--9$\mum$
complex and the 10.5--13.5$\mum$ complex
could be complicated by the 6.2, 7.7, 8.6,
11.3 and 12.7$\mum$ UIE bands,
the detection of the 3.26 and 4.69$\mum$ bands,
and, to a less degree, the 21 or 22$\mum$ band,
could {\it potentially} allow one to identify
cyanonaphthalenes in space.
%
%
Prior to JWST, the detection and identification of
the spectral features as possibly due to cyanonaphthalenes
would have been hampered by their small intensities
which would put them at the limit of modern
observational techniques,
including the {\it Short Wavelength Spectrometer} (SWS)
on board the {\it Infrared Space Observatory} (ISO)
and the {\it Infrared Camera} (IRC) on board {\it AKARI}.
With the advent of JWST, this will change.
The {\it Near InfraRed Spectrograph} (NIRSpec)
and {\it Mid-Infrared Instrument} (MIRI) 
on {\it JWST} span the wavelength range
of the characteristic vibrational bands
of cyanonaphthalenes.
%
{\it JWST}'s unique high sensitivity and
high resolution capabilities will open up
an IR window unexplored by
{\it Spitzer}\footnote{%
  The {\it Infrared Spectrograph} (IRS)
  on {\it Spitzer} only operates longward of
  $\simali$5.2$\mum$ and does not
  cover the characteristic 3.26$\mum$ C--H
  and 4.69$\mum$ C--N stretching bands
  of cyanonaphthalenes.
  }
and unmatched by {\it ISO} observations
and thus could potentially place
the detection of the IR vibrational bands
of individual PAH molecules on firm ground.

Admittedly, it is not clear if the detection
of the 3.26$\mum$ C--H and
4.69$\mum$ C--N stretching bands
could {\it uniquely} pinpoint the presence
of cyanonaphthalenes,
as other cyano-substituted PAHs
may also emit at similar wavelengths. 
Note that cyano-benzene (benzonitrile)
and cyano-indene have also been detected
in TMC-1 (McGuire et al.\ 2018, Sita et al.\ 2022).
To this end, a systematic calculation of
the IR emission spectra of these molecules
would be crucial. One can imagine that
different-sized cyano-containing aromatic molecules
would exhibit different C--H/C--N band ratios
since, with different energy contents,
they are expected to be excited to different
energy levels by the same stellar photon.

Let $\NH$ be the hydrogen column density
along the line of sight to an astrophysical region. 
Let $\left[{\rm C/H}\right]_{\rm CNN}$
be the number of carbon atoms (per H nucleon)
locked up in a cyanonaphthalene molecule.
The intensity of the IR emission
(erg$\s^{-1}\cm^{-3}\sr^{-1}$)
expected from cyanonaphthalenes would be
\begin{equation}
I_\lambda = j_\lambda\times\left(\frac{\NH}{11}\right)\times\Ccnn  ~~,
\end{equation}
where the denominator ``11'' accounts for the fact
that a cyanonaphthalene molecule has 11 carbon atoms.
The column density of cyanonaphthalenes
is simply
\begin{equation}
  N_{\rm CNN} = \left(\frac{\NH}{11}\right)\times\Ccnn  ~~.
\end{equation}
Therefore, by comparing the observed intensity
$I_\lambda^{\rm obs}$ with the model emissivity
$j_\lambda$ (e.g., see Figure~\ref{fig:Teff}a--d),
one can derive $N_{\rm CNN}$. If $\NH$ is known,
then one can determine how much carbon
is locked up in cyanonaphthalenes.
On the other hand, if $N_{\rm CNN}$ is known
(e.g., for 1-CNN or 2-CNN in TMC-1 from
the measurements of their rotational lines),
one can predict the IR emission intensity
$I_\lambda$ by multiplying the model emissivity
shown in Figure~\ref{fig:Teff}a--d with $N_{\rm CNN}$.

While cyanonaphthalenes are present
in the UV-attenuated TMC-1 molecular cloud,
it is not clear if they can survive in hostile regions
where the UV radiation is intense
(e.g., the diffuse ISM, the Orion Bar
and the M17 star-forming cloud).
According to Jacovella et al.\ (2022),
HCN removal from protonated benzonitrile
requires only $\simali$3.4$\eV$.
As similar HCN removal energies are
expected for cyanonaphtalenes,
the survivability of these species
in UV-irradiated environments
such as the Orion Bar is questionable.
It has been argued that only PAHs
with more than 20 carbon atoms
may survive in these regions (see Tielens 2008).
However, Stockett et al.\ (2023) recently found that
cyanonaphthalene can be efficiently stabilized
following ionization, with the aid of the so-called
``Recurrent Fluorescence''
(also known as Poincar\'e fluorescence,
see L\'eger et al.\ 1988),
a radiative relaxation channel
in which optical photons are emitted
from thermally populated electronically excited states.
Iida et al.\ (2022) also found that the Recurrent
Fluorescence could efficiently stabilize
small cationic carbon clusters
with as few as nine carbon atoms.
The Recurrent Fluorescence of PAHs
excited by UV photons could reach high
quantum efficiencies in photon conversion
and has been suggested
as the source of the so-called ``extended
red emission'' (ERE; Witt \& Lai 2020).

If the Recurrent Fluorescence does occur,
the absorbed stellar photon energy
will not be completely released
through vibrational transitions.
Therefore, the IR vibrational emissivity
is expected to be somewhat lower than
that calculated in Figure~\ref{fig:Teff}a--d,
roughly by a factor of
$\langle h\nu\rangle_{\rm RF}/\langle h\nu\rangle_{\rm abs}$,
where $\langle h\nu\rangle_{\rm RF}$ is the mean
energy of the optical photons emitted
due to Recurrent Fluorescence,
and $\langle h\nu\rangle_{\rm abs}$
is the mean enegy of the absorbed stellar photons.
For cyanonaphthalenes in the diffuse ISM,
the reflection nebula NGC\,7023,
and the Orion Bar,
$\langle h\nu\rangle_{\rm abs}$ is in the order
of 8--9$\eV$.
With $\langle h\nu\rangle_{\rm RF}\approx 2\eV$
(see Stockett et al.\ 2023),
the IR emissivity $j_\lambda$ will only be reduced
by $\simali$20\% and the emission spectral profile
will largely remain the same. 

Finally, we note that the UV absorption cross sections
of cyanonaphthalenes (i.e., 1-CNN and 2-CNN)
adopted in this work were ``synthesized'' from
1-CNN and naphthalene (see \S\ref{sec:cabs}).
It is not clear how different the actual UV absorption
cross sections of 1-CNN would be
compared with naphthalene
and of 2-CNN compared with 1-CNN.
As a priori, one would imagine that
the incorporation of a CN group
in naphthalene, the parental molecule
of cyanonaphthalenes, would cause
significant changes to the UV absorption spectrum,
as the $\pi$ electrons of the CN group are partially
conjugated with the main aromatic $\pi$ system.
Nevertheless, as mentioned earlier 
(see  Figure~\ref{fig:Teff}a--d),
the IR emission spectra of cyanonaphthalenes
are not sensitive to the illuminating starlight spectrum.
This also implies that the IR emission spectral shape
of cyanonaphthalenes is not very sensitive to the exact
UV absorption. In any case,  future experimental
measurements of the absorption spectra of
cyanonaphthalenes from the optical to the far-UV
will be helpful in examining their vibrational excitation.
%

We also note that the IR vibrational frequencies
and intensities of cyanonaphthalenes adopted
here were computed by Bauschlicher (1998),
using a DFT force field and intensitie
in the harmonic limit.
It is well recognized that anharmonic interactions
could appreciably affect the vibrational
frequencies and intensities.
In addition, Fermi interactions can have
a significant impact on the C--H stretches.
Moreover, quadratic dipole dependence
can also affect the vibrational intensities.
While some of these effects could be grossly
accounted for by empirical scaling coefficients
as applied by Bauschlicher (1998),
more accurate results may be obtained with
the standard second-order vibrational perturbation
theory (VPT2). Note that the VPT2 theory is feasible
with DFT force fields for molecules
as large as cyanonaphthalenes.
Future quantum chemical computations
utilizing the VPT2 theory and even better,
experimental measurements of gas-phase
cyanonaphthalenes will be helpful in evaluating
the accuracy of the IR emission spectra calculated here. 

\begin{figure*}
\begin{center}
\hspace{-1cm}
\begin{minipage}[t]{0.4\textwidth}
\resizebox{8.5cm}{7.5cm}{\includegraphics[clip]{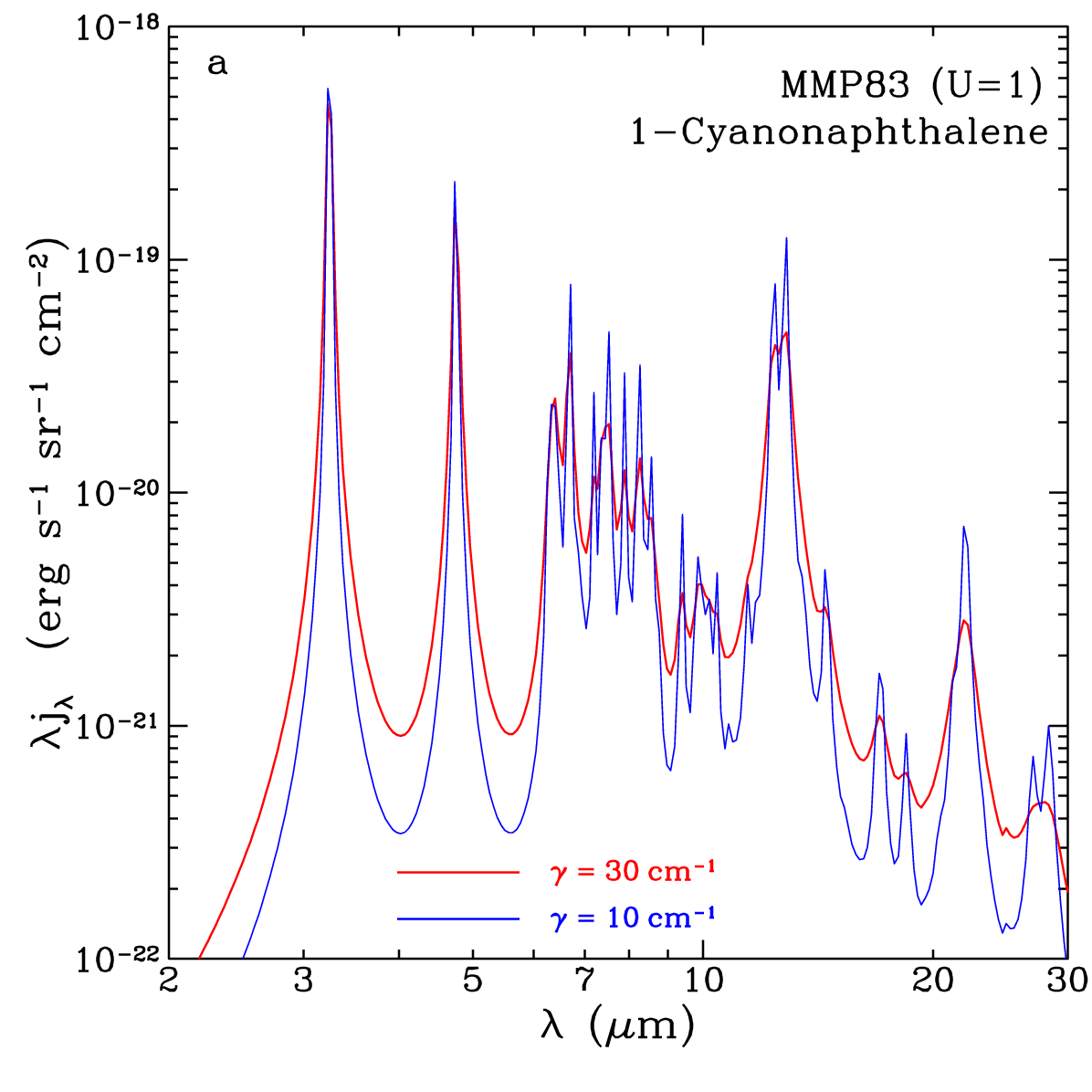}}\vspace{-0.5cm}
\end{minipage}
\hspace{2cm}
\begin{minipage}[t]{0.4\textwidth}
\resizebox{8.5cm}{7.5cm}{\includegraphics[clip]{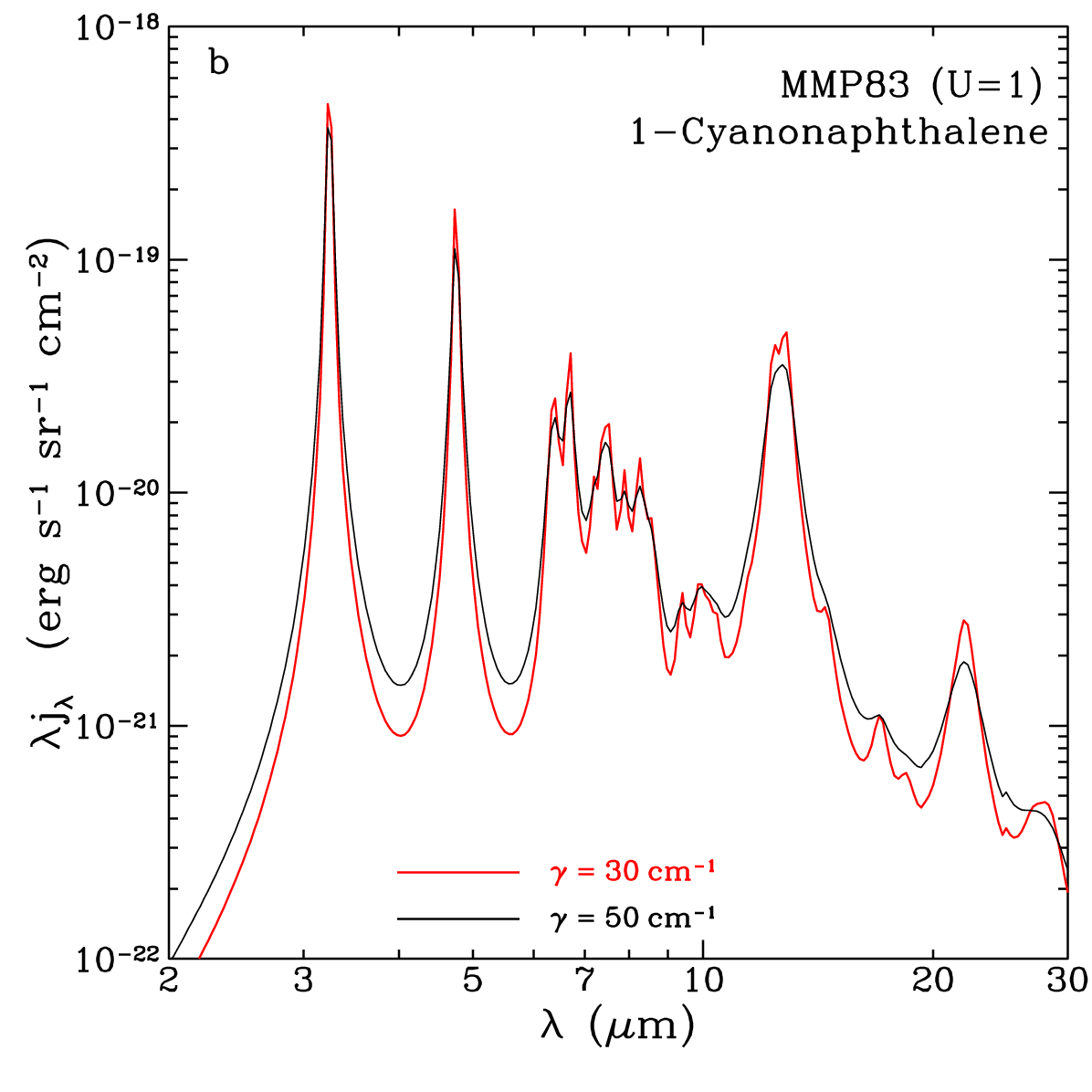}}\vspace{-0.5cm}
\end{minipage}
\end{center}
\caption{\footnotesize
         \label{fig:gamma}
         Left panel (a): Comparison of the IR emission spectra
         of 1-CNN of a natural line width of
         $\gamma=30\cm^{-1}$ (red line)
         for each vibrational transition
         with that of $\gamma=10\cm^{-1}$ (blue line).
         The molecule is illuminated by
         the MMP83 ($U$\,=\,1)
         interstellar radiation field. 
         Right panel (b): Same as (a) but for
         a comparison of $\gamma=30\cm^{-1}$ (red line)
         with  $\gamma=50\cm^{-1}$ (blue line).
       }
\end{figure*}

In calculating the IR emission sepctra of
cyanonaphthalenes, we assume a natural
line width of $\gamma=30\cm^{-1}$
for each vibrational transition
(see \S\ref{sec:cabs})
due to internal vibration redistribution (IVR).
This may be somewhat simplified.
Molecules like cyanonaphthalenes
are in a somewhat intermediate size-range,
where IVR is prevalent, but not necessarily
ergodic or completely statistical.
For these molecules,
spectral broadening may more likely
be dominated by anharmonic shifts of
the thermally excited vibrational levels.
Indeed, Hewett et al.\ (1994) argued that
the C--H fundamentals of naphthalene
are far from the IVR threshold.
In view of these, we have also considered
$\gamma=10\cm^{-1}$ and $50\cm^{-1}$.
As shown in Figure~\ref{fig:gamma},
the model IR emission spectra are essentially
the same, particularly in the characteristic
C--H and C--N stretching regions. 
This justifies the choice of a natural
line width of $\gamma=30\cm^{-1}$.

\section{Summary}\label{sec:summary}
We have modeled the vibrational excitation of
cyanonaphthalenes, the first specific PAH species
ever identified in space, and have calculated their
IR emission spectra for a number of representative
astrophysical regions, from the diffuse ISM to regions
illuminated by stars of different effective temperatures.
Also calculated includes the TMC-1 dark molecular
cloud where cyanonaphthalenes were first detected
through their rotational lines. It is found that while
the emissivity level varies from one region to another,
the overall IR emission spectra do not vary much with
environments or between two isomers of cyanonaphthalene
(i.e., 1-CNN and 2-CNN). Both isomers exhibit prominent
bands at 3.26$\mum$ (C--H stretch),
4.69$\mum$ (C--N stretch),
21 or 22$\mum$ (skeletal bend),
and broad complexes at $\simali$6--9$\mum$
and $\simali$10.5--13.5$\mum$.
The NIRSpec and MIRI instruments on board JWST
are well suited for searching for cyanonaphthalenes
in space, particularly through the characteristic bands
at 3.26 and 4.69$\mum$.

\acknowledgments{%
We thank the anonymous referees
for helpful comments and suggestions.
We thank B.M.~Broderick, B.T.~Draine,
B.A.~McGuire, and E.F.~van Dishoeck
for stimulating discussions.
KJL and TTF are supported by
the National Key R\&D Program of China
under No.\,2017YFA0402600,
and the NSFC grants 11890692,
12133008, and 12221003,
as well as CMS-CSST-2021-A04.
XJY is supported in part by
NSFC~12333005 and 12122302
and CMS-CSST-2021-A09.
}


\end{document}